%% file: main.tex
\documentclass[conference]{IEEEtran}

\IEEEoverridecommandlockouts
\usepackage[noadjust]{cite}
\usepackage{amsmath,amssymb,amsfonts}
\usepackage{glossaries}

\usepackage{graphicx}
\usepackage{textcomp}
\usepackage{xcolor}
\usepackage[capitalise]{cleveref}
\usepackage{siunitx}
\usepackage{tabularx}
\usepackage{colortbl}
\usepackage{algpseudocode}
\usepackage{algorithm}
\usepackage{xurl}
\usepackage{flushend}
\usepackage{tablefootnote}
\usepackage{bm}
\usepackage{booktabs}
\usepackage{makecell}   

\usepackage[export]{adjustbox}
\usepackage[caption=false,font=normalsize,labelfont=sf,textfont=sf]{subfig}
\usepackage[raggedrightboxes]{ragged2e}

\DeclareSIUnit[]\gforce{\mathcal{g}}

\crefname{subsection}{Subsection}{Subsections}

\DeclareSIUnit\operations{op}
\DeclareSIUnit\gforce{g}

\usepackage[colorinlistoftodos,prependcaption,textsize=tiny]{todonotes}

\def\BibTeX{{\rm B\kern-.05em{\sc i\kern-.025em b}\kern-.08em
    T\kern-.1667em\lower.7ex\hbox{E}\kern-.125emX}}

\glsdisablehyper 
\input{acronyms.tex}

\begin{document}

\title{UnReference: analysis of the effect of spoofing on RTK reference stations for connected rovers}

\author{
	\IEEEauthorblockN{
        Marco Spanghero\IEEEauthorrefmark{1} and Panos Papadimitratos\IEEEauthorrefmark{1},~\IEEEmembership{Fellow,~IEEE}
        }
        \\
	\IEEEauthorblockA{ \IEEEauthorrefmark{1}Networked Systems Security (NSS) Group -- KTH Royal Institute of Technology, Stockholm, Sweden \\
        marcosp@kth.se, papadim@kth.se}
}

\maketitle

\begin{abstract}
Global Navigation Satellite Systems (GNSS) provide standalone precise navigation for a wide gamut of applications. Nevertheless, applications or systems such as unmanned vehicles (aerial or ground vehicles and surface vessels) generally require a much higher level of accuracy than those provided by standalone receivers. The most effective and economical way of achieving centimeter-level accuracy is to rely on corrections provided by fixed \emph{reference station} receivers to improve the satellite ranging measurements. Differential GNSS (DGNSS) and Real Time Kinematics (RTK) provide centimeter-level accuracy by distributing online correction streams to connected nearby mobile receivers typically termed \emph{rovers}.
However, due to their static nature, reference stations are prime targets for GNSS attacks, both simplistic jamming and advanced spoofing, with different levels of adversarial control and complexity. Jamming the reference station would deny corrections and thus accuracy to the rovers. Spoofing the reference station would force it to distribute misleading corrections. As a result, all connected rovers using those corrections will be equally influenced by the adversary independently of their actual trajectory. 
We evaluate a battery of tests generated with an RF simulator to test the robustness of a common DGNSS/RTK processing library and receivers. We test both jamming and synchronized spoofing to demonstrate that adversarial action on the rover using reference spoofing is both effective and convenient from an adversarial perspective. Additionally, we discuss possible strategies based on existing countermeasures (self-validation of the PNT solution and monitoring of own clock drift) that the rover and the reference station can adopt to avoid using or distributing bogus corrections. 

\end{abstract}

\section{Introduction}
\label{sec:introduction}
\gls{gnss} services are commonly used to provide precise timing and localization, globally, to a wide set of applications. The evolution of \gls{gnss} systems with additional constellations and frequencies, with interoperability, has led to sub-meter accuracy in mass production commercial devices; even without relying on advanced antenna systems. Nevertheless, applications requiring highly precise localization, such as topographic surveys, precision navigation, and machine control, are still dependent on external corrections based on \gls{rtk}. In such setting, a moving receiver (rover) corrects its observation based on measurements of the same satellites provided by a reference station with a known fixed position. Usually, such corrections complement and improve the ones provided by multi-frequency systems and allow the rover, after a usually fast convergence period, to achieve a \gls{pnt} solution with 
an accuracy of a few centimeters. Multiple correction delivery methods exist, relying on terrestrial infrastructure, network connectivity, or, more recently, satellite L-band downlink. 

This work focuses on corrections delivered over the Internet. Specifically, network-based \gls{rtk} (NRTK) is established as an open-source, collaborative system to achieve centimeter-level accuracy with consumer-grade receivers \cite{ntrip_esa}. For this reason, we believe the majority of the nonindustrial market will rely on NRTK for a long time before a complete adoption of the L-Band correction services, given NRTK large availability even free of charge for the final user thanks to several open source projects \cite{Dabove2023,snip}. While the up-and-coming L-Band correction streams are potentially more secure (due to the proprietary and often encrypted signal structure), to the best of our knowledge they are still quite specialized and cannot be expected to be free of charge for the final user. Examples of L-Band correction services cover both legacy signals and external corrections delivered over Iridium and L-band (\cite{fugroAtomichron,spartn}) or more modern approaches such as Galileo High Accuracy Services (HAS) in the E6 band. 

\gls{rtk} reference stations are mounted at precisely surveyed points, whose location is accurately determined and publicly available. This way, the reference receiver can calculate precise carrier phase measurements referenced to its static position and distribute this information to the connected rovers which ultimately use them to calculate precise positions. The corrections are meaningful only if the rover can securely access the reference station (e.g., via an authenticated and encrypted channel, \cite{Pepjin:2020}) and the reference station itself is not under adversarial control. 

Given their intrinsically static placement and the current unencrypted and public nature of the \gls{gnss} signals make RTK stations a relatively easy target for manipulation, via spoofing, meaconing (e.g. replay/relay), or jamming \cite{HumphreysAssessingSpoofer, LenhartSP:C:2022, 10.1145/3558482.3590186}. 
Although cryptographic countermeasures thwarting spoofing exist, adopting such methods will take time especially the signal in space or the receiver structure \cite{PapadimitratosJa:C:2008} require modifications. OSNMA (\cite{8714151,Gotzelmannnavi.572}) stops the attacker from simulating \gls{gnss} signals with invalid information, but remains susceptible to meaconing as there is no cryptographical protection in the spreading code itself (\cite{ZhangP:C:2019a,Wang2023}). 

Spoofing any mobile GNSS receiver is complex, while it is much simpler and straightforward to smoothly capture a GNSS receiver whose location is well-determined and fixed, especially if the attacker gains line of sight to the victim's antenna. The attacker can target and significantly degrade the \gls{pnt} solution at the \gls{rtk} reference station. This in turn will directly produce incorrect corrections conveyed to the rover, and, if such corrections are applied blindly, potentially result in highly degraded trajectory estimation or outright denial of service, as the \gls{rtk} algorithm might not converge to a viable solution. Practically, each \gls{rtk} reference receiver, even if otherwise trusted, can be jammed or spoofed: cryptographical protocols safeguarding the network-based correction distribution system cannot contain this problem. 

In this work, we analyze and expand promising preliminary results that show how the rover \gls{pnt} quality is affected by a degraded reference station. For different \gls{gnss} constellation configurations and attack methods, we show that a strategically placed attacker can degrade the \gls{pnt} solution quality at the rover simply by attacking the reference station receiver. We demonstrate three adversarial settings (synchronous single constellation lift-off spoofing, multiconstellation asynchronous spoofing, and jamming), and the effects they have on the rover \gls{pnt} solution.

The rest of the paper is organized as follows: \cref{sec:sys} describe the system model, focusing on the aspects relevant to the \gls{rtk} problem, \cref{sec:methodology} discusses adversary model and control approach, \cref{sec:experimental} presents our experimental setting, the devices used, and the different adversarial scenarios. \cref{sec:evaluation} discusses our findings before we conclude in \cref{sec:conclusion}, highlighting possible future directions for improving the security of \gls{rtk} augmentation networks. 
 
\section{Related Work}
\label{sec:related_work}
\gls{rtk} correction services are often assumed to be trusted. Few works investigated the effects of spoofing and jamming the reference stations on the connected rovers. 
Empirical evaluation of the effect of jamming on a reference station was shown in \cite{gpspatron}, based on live observations. Similarly, \cite{6106252} analyzes only the DGNSS case, considering a limited case of code-shift spoofing on a few selected signals. 

A network adversary injecting fake correction information or impersonating a reference station to degrade the rover's \gls{pnt} solution is considered in \cite{Pepjin:2020}. While this is potentially problematic, such vulnerability can be addressed with secure networking between the rover and reference station. For example, \gls{tls} based methods \cite{rfc5246} can authenticate and protect the integrity and confidentiality of the data sent to the rover, relying on established public key cryptography at the cost of increased complexity for the infrastructure and the credential management system. However,  albeit fundamental and needed, network security cannot mitigate adversaries targeting directly the  reference station  \gls{gnss} receiver by controlling the received \gls{gnss} signals.

\cite{HumphreysAssessingSpoofer,SpangheroPP:C:2023,Gao2020,Blum2021,tippenhauer2011requirements} demonstrated how even advanced receivers are generally sensitive to adversarial interference either in the form of jamming (targeting an outright \gls{gnss} denial of service at the victim) or spoofing (aiming at providing the \gls{gnss} receiver with valid but manufactured) \gls{gnss} signals. Jamming has been effective in degrading the receiver's solution up to the level of outright unavailability, and it is relatively simple to carry out as it does not require any complex hardware or sophisticated technique. 
Reference stations are generally designed to be resilient to jamming by using specialized design antennas (choke ring antennas to limit the effectiveness of jammers placed on the ground) or high-performance adaptive filters (adaptive notch filters to suppress even fast narrow band signals). Additionally, jamming/spoofing detection and localization systems can help in increasing the efficiency of the removal process even over large areas, where the jammer might be located \cite{SpangheroMGPP:C:2024}. 

Differential \gls{gnss} and carrier estimation can benefit antispoofing, but the reference station information is considered to be valid and authentic (\cite{WOS:000356331204003,278b52b90206426bae50f2f582b4aeb1,WOS:000375213003001,Hu_Bian_Ji_Li_2018}). Methods (usually applied to static receivers, like the case of time servers) provide a layer robustness from spoofers by analyzing both the movement of the antenna (to detect changes not consistent with a static reference), but mostly focus on the quality of the provided time transfer. 

\gls{sqm} techniques that track changes in code/doppler and $C/N_0$ also contribute positively to the security of the reference station (\cite{Spensnavi.537,10081330,WOS:000305396000003,akos2003,WOS:000209006500003,WOS:000429990900016}). Specifically, interference monitoring based on \gls{agc} and $C/N_0$ can work well given the static nature of the station and the fact that generally such reference points are set in low multipath environments. Still, it is possible for an adversary to precisely match the $C/N_0$ for each satellite at the victim receiver and transmit signals that are slightly above that, to avoid triggering detection by means of power envelope monitoring. On the other hand, the signals can be carefully aligned to the real one in matters of power, code, and carrier phase offset \cite{HumphreysAssessingSpoofer,SpangheroPP:C:2023}. Even in the case of time servers, synchronous spoofing proved capable of modifying the \gls{pnt} solution of the receiver. 

Adversarial techniques allowing progressive stretching of all pseudoranges for the satellites in view (which do not modify the position of the victim receiver) are potentially critical for reference stations as they modify the pseudorange information of the carrier phase corrections \cite{SpangheroPP:C:2023,Jiadong2019}. This is particularly problematic for differential \gls{gnss} which relies on single differencing with the reference station to achieve increased precision. \gls{rtk} relies on carrier phase measurements to achieve higher accuracy, but still, this means that to achieve a full \gls{rtk} fix the carrier phase measured at the reference station needs to be coherent with its position and measured pseudoranges.

Additionally, meaconing can also effectively control a victim \gls{gnss} receiver, even if at a much lower level of flexibility compared to simulation-based spoofing. Signal level meaconing is problematic for cryptographically enhanced signals as it replicates the signal identically, allowing the attacker to bypass any authentication mechanism, and select specific bands by filtering at the meaconer which channels to relay. 

\section{System model}
\label{sec:sys}

\begin{figure}[]
    \centering
    \includegraphics[width=.8\linewidth]{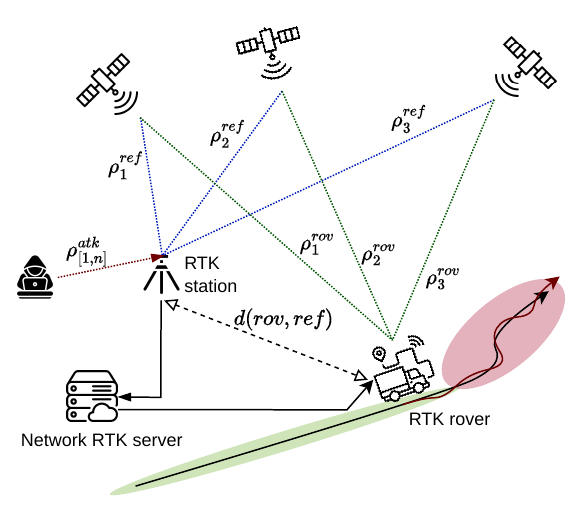}
    \caption{Typical network \gls{rtk} scenario with an attacker spoofing the reference station}
    \label{fig:atk-pictorial}
\end{figure}

Two \gls{gnss} receivers, one configured as \gls{rtk} rover and the other as \gls{rtk} station, are connected over a network link. The \gls{rtk} station is at a fixed point with known coordinates, included in the correction stream provided to the rovers over the network. The network link between the rover and the station is protected by standard network security (e.g., authenticated link) whose deployment is beyond the scope of this work. The adversary can attack the RTK reference station \gls{gnss} receiver by jamming, simulation, co-simulation, or replay-relay of \gls{gnss} signals. The attack setup is shown in \cref{fig:atk-pictorial}: the attacker causes a reduction of accuracy at the rover by the transmission of interference to the reference station.

It is not important in the scope of this work how many constellations and frequency bands the adversary can control via spoofing, as this only influences the requirements for its computational power. It is always possible for the adversary to jam the signals it is not spoofing, to block the receiver from obtaining legitimate information from those channels. 

\begin{figure*}
    \centering
    \includegraphics[width=0.8\textwidth]{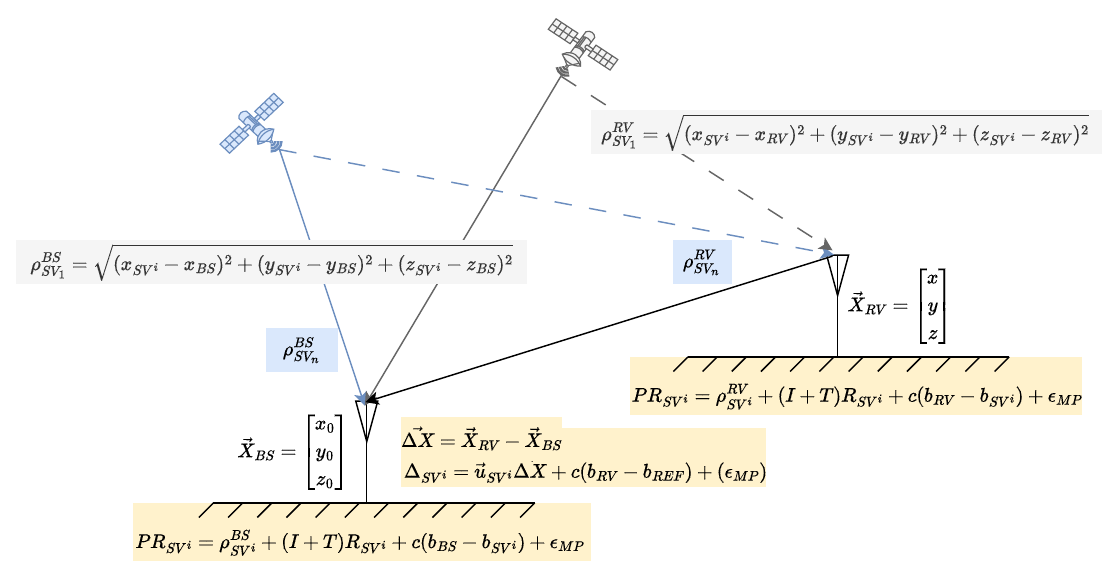}
    \caption{Single differencing in DGNSS. The corrections are valid for the same observation pairs at the rover and the reference stations and usually up to \SI{100}{\kilo\meter}.}
    \label{fig:dgnss-pictorial}
\end{figure*}

Differential \gls{gnss} and \gls{rtk} are two different approaches to augment the accuracy of a moving \gls{gnss} receiver. We will give here a brief background of each to show how an attacker can influence the reference station.
Traditional differential \gls{gnss} (DGNSS) operates at the pseudorange level, correcting the rover's observations by single-differencing them with the observations at the station. This process is shown in \cref{fig:dgnss-pictorial}.
The pseudorange for a generic satellite at a \gls{gnss} receiver is given in \cref{eq:pr_rx}, where $\rho$ is the geometrical range between satellite and receiver, $I,T$ are the ionospheric and tropospheric delays, $\epsilon_{MP}$ is the multipath induced error term.
\begin{equation}
    PR_{SV^i} = \rho_{SV^i} + (I+T)R_{SV^i} + c(b_{RX} - bR_{SV^i}) + \epsilon_{MP}
    \label{eq:pr_rx}
\end{equation}
The geometrical range is a function of the actual receiver position on the ground, as in \cref{eq:range_rover} and \cref{eq:range_ref}, for the rover and the reference station respectively
\begin{equation}
    \rho^{RV}_{SV^i} = \sqrt{(x_{SV^i}-x_{RV})^2+(y_{SV^i}-y_{RV})^2+(z_{SV^i}-z_{RV})^2}
     \label{eq:range_rover}
\end{equation}
\begin{equation}
    \rho^{BS}_{SV^i} = \sqrt{(x_{SV^i}-x_{BS})^2+(y_{SV^i}-y_{BS})^2+(z_{SV^i}-z_{BS})^2}
     \label{eq:range_ref}
\end{equation}

While the location of the rover is obviously unknown, the location of the reference station is well defined, usually with a long-term survey-in process which consists of averaging the observations and the \gls{pnt} solution at the reference receiver for long periods of time (\SI{24}{\hour} is generally considered sufficient, or until convergence to the required accuracy is reached). 

By a single difference of pseudoranges, we express the position of the receiver as a function of the unit vector from the reference to each satellite in view and the distance between the rover and the reference station. In \cref{eq:single_diff}, the corrections for the rover are computed by eliminating propagation effects and assuming a single multipath term.  

\begin{equation}
    \Delta_{SV^i} = \bm{u}_{SV^i}\Delta\bm{x} + c(b_{RV} - b_{REF}) + (\epsilon_{MP})
    \label{eq:single_diff}
\end{equation}

The rover applies the $\Delta_{SV^i}$ correction per satellite (\cref{eq:single_diff}) and uses the corrected observations to calculate its position iteratively, in respect of the reference station. 

The more modern approach to differential corrections builds upon DGNSS but adds a second differencing step to also account for the carrier phase term. This allows the rover \gls{gnss} receiver to achieve centimeter-level accuracy but at a higher processing cost. We express the carrier phase at the rover and the reference station as shown in \cref{eq:carrier}, where we omit the multipath and atmospheric propagation terms which are the same as in \cref{eq:pr_rx}

\begin{equation}
    \Phi^{RX}_{SV_i} = \frac{2\pi}{\lambda}(\rho^{RX}_{SV_i} - N^{Rx}_{SV_i}\lambda + c(b_{RX} - bR_{SV^i}))
    \label{eq:carrier}
\end{equation}

\begin{figure*}
    \centering
    \includegraphics[width=0.8\textwidth]{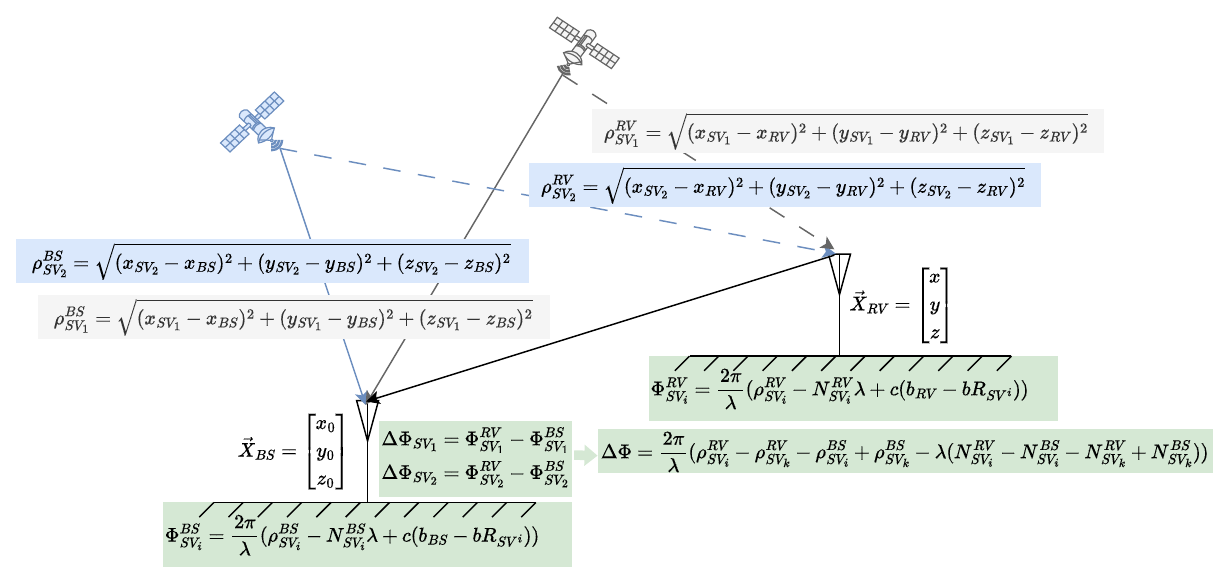}
    \caption{Double differencing in \gls{rtk}. The corrections are valid for the same observation pairs at the rover and the reference stations and usually up to \SI{2}{\kilo\meter}.}
    \label{fig:rtk-pictorial}
\end{figure*}

The double differencing process adds an additional step following \cref{eq:single_diff}. The second differencing removes the propagation errors between the satellites themselves, as it eliminates the biases introduced by the satellite and receiver clock biases, as shown in \cref{eq:doublediff_1,eq:doublediff_2} and in \cref{fig:rtk-pictorial}.

\begin{equation}
    \left\{
    \begin{aligned}
    \Delta\Phi_{SV_i} &= \Phi^{RV}_{SV_i} - \Phi^{BS}_{SV_i} \\ 
    \Delta\Phi_{SV_k} &= \Phi^{RV}_{SV_k} - \Phi^{BS}_{SV_k} \text{  , for $k\neq i$}
    \end{aligned}
    \right.
    \label{eq:doublediff_1}
\end{equation}

\begin{equation}
    \begin{aligned}
        \Delta\Phi & = \frac{2\pi}{\lambda}(\rho^{RV}_{SV_i} - \rho^{RV}_{SV_k} - \rho^{BS}_{SV_i} + \rho^{BS}_{SV_k} \\
        & - \lambda(N^{RV}_{SV_i} - N^{BS}_{SV_i} - N^{RV}_{SV_k} + N^{BS}_{SV_k}))
    \end{aligned}
    \label{eq:doublediff_2}
\end{equation}

\section{Attacker approach}
\label{sec:methodology}

Due to the open structure of the \gls{gnss} signal the adversary can craft signals with valid modulation, frequency, and data content that match the position of the \gls{rtk} station and, jointly, the adversarial action. This usually consists of (but is not limited to) modifications in the navigation data information, code-carrier modification, and selective replacement of specific signals \cite{HumphreysAssessingSpoofer,LenhartSP:C:2022}. Specifically, the adversary can match and synchronize the spoofing signals to the legitimate ones and slowly force its adversarial action. It is important to notice here that the attacker aims to disrupt and degrade the \gls{rtk} solution quality at the rover. In a regular spoofing scenario, the adversary cannot usually measure the effect of its action on the victim receiver. On the other hand, RTK corrections are typically publicly available - if so, the adversary can connect its rover and this way monitor the degradation caused by tampering with the reference station.

An adversary spoofing the reference stations can achieve its target by: (i) changing the pseudorange observations that the measured ranges do not match the real ones, (ii) modifying the clock offset of the reference station, (iii) modifying the atmospheric propagation terms to introduce an error in the corrections. 

In \cref{eq:single_diff} this effect is seen directly in the correction terms either as a change in the reference station clock bias or in the pseudoranges observations. Two aspects need to be taken into consideration at this point. The adversary cannot modify the clock solution beyond the validity of the corrections at the rover. Practically, this means that the adversary can change the RTK station clock offset, but cannot abruptly change the time solution at the references. Additionally, the attacker should avoid changing the reference station position while mounting the attack, to avoid simplistic detection based on position validation. Such deviations often trigger a restart of the reference station automatic establishment process (often called survey-in) which generally makes correction unavailable during the convergence period.

In \cref{eq:doublediff_1,eq:doublediff_2} the action of the attacker is similar but needs to be more surgical. As \gls{rtk} relies on carrier phase information, the attacker needs to modify the carrier phase per each signal while being consistent with induced change in the pseudoranges. This requires the adversary to shift the phase of the signals, either independently or altogether. This task is complex, but an attacker can mount a stealthy control action at the victim if the latter is static and its position is well-known and determined. 

Jamming is also known to affect the accuracy of the receiver, even without causing a direct denial of service. A reduction in the \gls{snr} is already sufficient to reduce the observation quality. The efficiency of the attack depends on the chosen jamming waveform, which can vary from simple blanket noise transmission to more complex chirp or sweeping patterns. While such an attack is less complex and requires a more simplistic approach, it has proven effective in degrading the quality of the receiver \gls{gnss} measurements.

\section{Experimental setup}
\label{sec:experimental}

Our system setup includes two u-Blox ZED-F9P multi-frequency, high-precision \gls{gnss} receivers, each connected to a platform capable of providing connectivity and computation. One device is configured to broadcast \gls{rtk} corrections over a secure channel using a standard \gls{ntrip} provider to all connected clients. For the purpose of this work, it is not important how the rover and the reference station exchange information in a secure, authenticated way (e.g., this can be implemented with secure network transport). 
The rover receiver provides raw \gls{gnss} measurement data to an implementation of RTKLib (open source, \cite{rtklibexplorer}) that processes the \gls{rtk} solution based on the correction stream obtained from the \gls{ntrip} server \cite{ntrip_esa}. Additionally, we test the internal \gls{rtk} engine of the u-Blox ZED-F9P, to compare its performance to RTKLib. This is relevant as often corrections are also distributed over local short-range wireless networks, especially in the case of semi-permanent installations as the ones often used for \gls{uav} guidance or construction work. 


The ground truth and the adversary are implemented using Safran Skydel, a fully featured tool to generate \gls{gnss} RF simulation scenarios \cite{skydel}.
To achieve control of the reference station the adversary can approach the problem from two different points of view. First, if no authentication on the navigation messages is present, the adversary crafts bogus navigation messages reporting fake orbital parameters for a specific satellite. This will cause the reference station to broadcast wrong ephemerides corrections, and calculate the wrong satellite position for the pseudorange and carrier corrections. Alternatively, the attacker can mount a more subtle attack, where the navigation messages are correct. Still, the code/carrier phases are shifted to make the pseudorange and carrier offset calculations incorrect for the specific reference station position. If authentication of the navigation messages is available, the adversary can choose to deny a particular installation relying on jamming or can resort to meaconing to replay signals sampled at a different location \cite{LenhartSP:C:2022} or rely on more complex signal generation techniques accounting for the cryptographic methods used \cite{ZhangLP:J:2022,9085417}.

\begin{figure*}
    \centering
    \includegraphics[width=\linewidth]{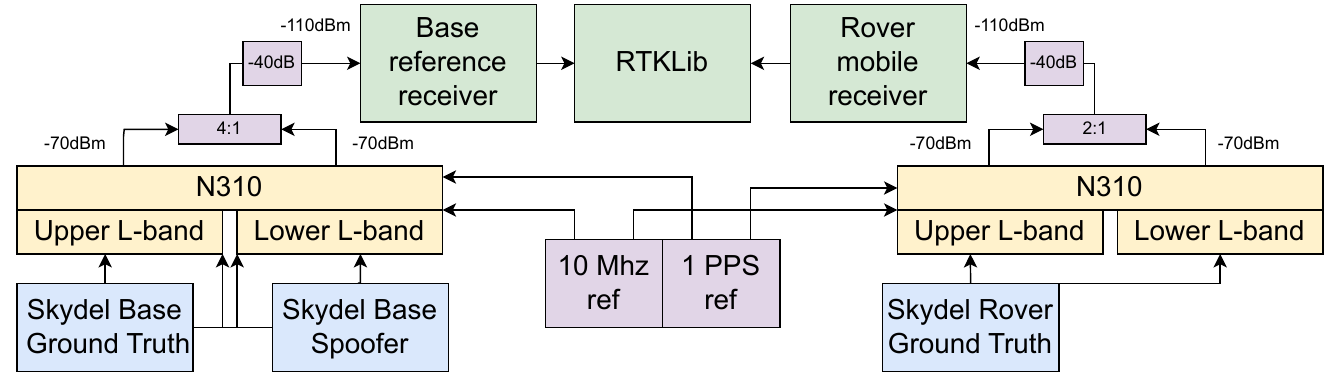}
    \caption{Experimental setup}
    \label{fig:experiment-pictorial}
\end{figure*}

Such attacks are evaluated along with their combination with jamming. In the synchronous attack, code-phase aligned signals are generated, in all constellations visible and configured at the receiver. Such signals are consistent with the \gls{rtk} station's live sky view and the transmit time is synchronous with the beginning of the GPS frame at the victim receiver. We also test unsynchronized spoofing, where the adversary overtakes the victim by overpowering the legitimate signals. Such attacks often do not require initial jamming, as the incoherent interference of the spoofing acts as jamming in the first phase, until the \gls{gnss} receiver shifts to the simulated signals. An overview of the experimental test setup is given in \cref{fig:experiment-pictorial}.

\begin{figure}[ht!]
    \centering
    \includegraphics[width=\linewidth]{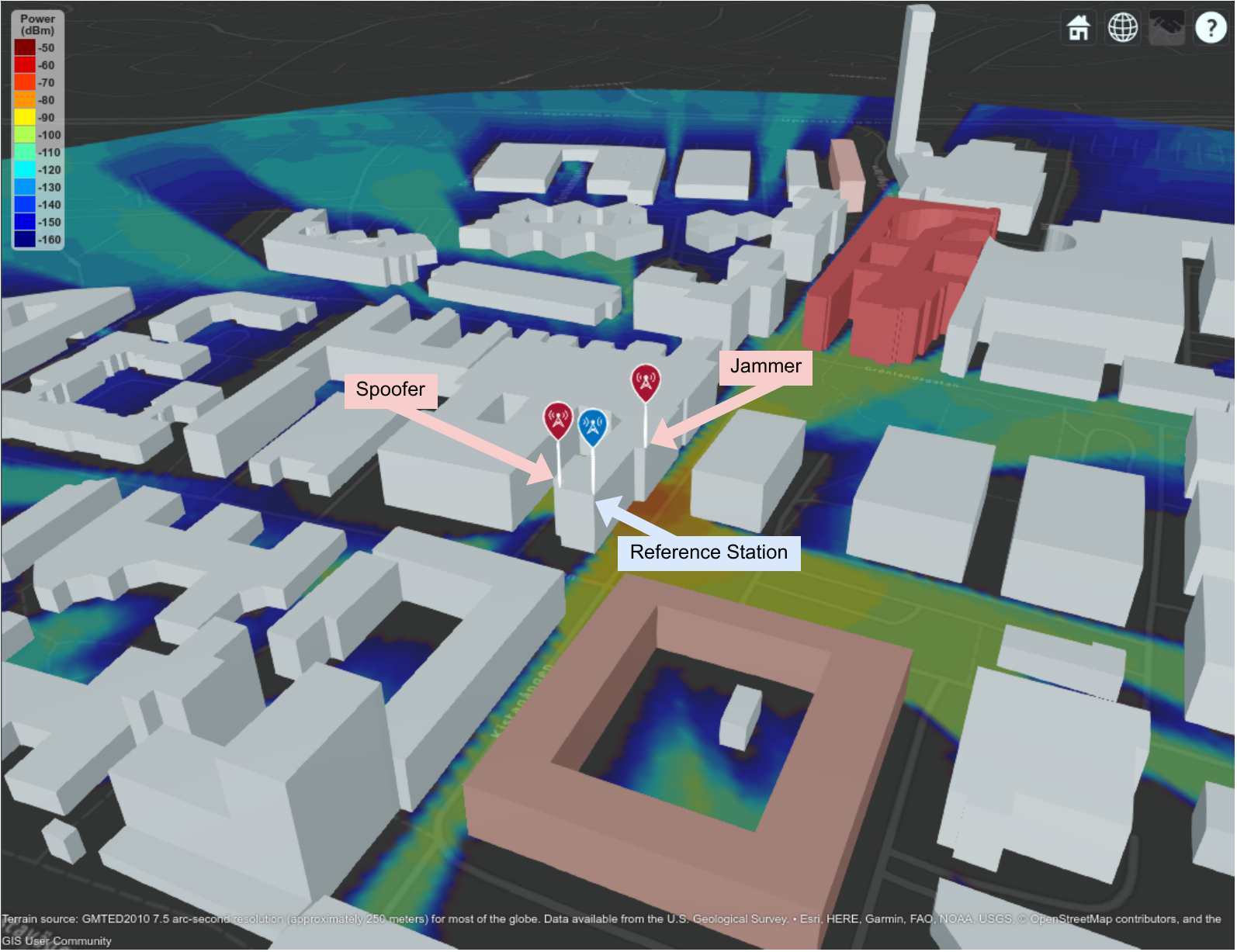}
    \caption{RF signal simulation setup and relative position of the spoofer and jammer to the reference station.}
    \label{fig:experiment-setup}
\end{figure}

The tests are conducted in a protected environment so as not to cause involuntary disturbance in the protected \gls{gnss} frequency bands. The simulation is at signal level, meaning that the generated signals take into account possible propagation issues and environment-induced variations, within the limits of what is available within the Safran Skydel software. \cref{fig:experiment-setup} shows the setup used for the RF simulations. The setup is a jammer and a spoofer, with the latter having LoS with the RTK reference station. In the case of the jammer, we assume an omnidirectional antenna, while the spoofer could be using a directive antenna. The overlay shows the modeled propagation of the jammer based on a Longley-Rice propagation model. The 3D model of the buildings is provided by Openstreetmaps data.

We perform a set of different tests, that include jamming, spoofing, or a combination of the two. The location of the reference station is fixed and corresponds to a real reference station that is publicly available and operated by the NSS Group at KTH \cite{nssgrouprtk}. The rover moves on a simulated trajectory and includes different types of environments. The mobility model used is the one of a vehicle and the trajectory is designed to have realistic dynamics and acceleration. The entire trajectory is contained in a radius of \SI{5}{\kilo\meter} from the reference station to guarantee the validity of the corrections.


\section{Results}
\label{sec:evaluation}

To evaluate the attack effect we construct a battery of tests that includes various combinations of attack, frequency used by the receiver, and constellations. We focus on Galileo and GPS, for the L1/L2 and E1/E5b respectively as these are supported by the F9P receivers that we used in our setup. The coordinate system used is either WGS84 (for the mapping) or ECEF (for the error measurements). The tests and corresponding \gls{rtk} configurations are described in \cref{tab:test_setup} and summarized as follows:
\begin{itemize}
    \item Overpowered synchronous multi-constellation spoofing with single frequency spoofer: Tests 1.(a,b,c);
    \item Single frequency jammer: Tests 2.(a,b);
    \item Single frequency, single constellation spoofer: Test 3.(a,b);
    \item Commercial implementation, multi-constellation and multi-frequency spoofer: Test 4.a
\end{itemize}
The RTKLib implementation used is unable to operate in stand-alone mode without corrections if \gls{rtk} is required and the jamming signal is designed to test the rover's behavior to the degraded solution at the reference station but without actual denial of service. 

\textbf{Test configurations -} Tests 0.(a,b,c) are used as ground truth for the other cases, and different combinations of signals at the rover are used to test both the accuracy and quality of the solution. Tests~1.(a,b) are designed to test the robustness of a multi-constellation/multi-frequency receiver to a single-frequency spoofer, operating without a jammer. Test~1.(c), considers a multi-constellation single-frequency reference station and rover only. Tests~2.(a,b) evaluate the effects of jamming on a reference station for a single-frequency single-constellation reference station where the rover is a multi-frequency/constellation receiver. Tests 3.(a,b) instead assume a single-frequency and constellation reference station serving a multi-frequency constellation receiver with a selective spoofer. In this case, the spoofer only controls the L1 GPS signals, leaving the Galileo signals unmodified. This allows us to test the selectiveness of the attack and the performance of each system. Notably, the L1 spoofer acts as a BPSK modulated jammer for the E1 signal, but with less power budget due to the matched transmission power to be an effective spoofer. Test 4.a considers a more powerful attack capable of controlling the full multi-constellation, multi-frequency receiver. 

\begin{table*}
\centering
\caption{RTK tests performed in the RF level simulation. All tests are conducted against physical receivers by simulating realistic signals.}
\label{tab:test_setup}
\resizebox{\textwidth}{!}{
\begin{tabular}{@{}p{0.8in} p{1in} p{1.3in} p{1.3in} p{1.1in} >{\raggedleft\arraybackslash}p{1in} >{\raggedleft\arraybackslash}p{1.7in}@{}}
    \toprule
    \textbf{Test} & \textbf{Trajectory} & \textbf{Reference Bands} & \textbf{Rover Bands} & \textbf{Jamming} & \textbf{Spoofing} & \textbf{RTCM provider} \\
    \midrule
    Test 0.a & Kista\_driving & L1/E1 + L2/E5b & L1/E1 + L2/E5b & No & No & Station Receiver (unspoofed) \\
    Test 0.b & Kista\_driving & L1/E1 + L2/E5b & L1 + L2 & No & No & Station Receiver (unspoofed) \\
    Test 0.c & Kista\_driving & L1/E1 + L2/E5b & E1 + E5b & No & No & Station Receiver (unspoofed) \\
    \addlinespace
    \midrule
    Test 1.a & Kista\_driving & L1/E1 + L2/E5b & L1/E1 + L2/E5b & No & \makecell{Synchronous, L1/E1} & True simulated (unspoofed) \\
    Test 1.b & Kista\_driving & L1/E1 + L2/E5b & L1/E1 + L2/E5b & No & \makecell{Synchronous, L1/E1} & Station Receiver (spoofed) \\
    Test 1.c & Kista\_driving & L1/E1 & L1/E1 & No & \makecell{Synchronous, L1/E1} & Station Receiver (spoofed) \\
    \addlinespace
    \midrule
    Test 2.a & Kista\_driving & L1 & L1/E1 + L2/E5b & \makecell{BPSK, \SI{25}{\mega\hertz}, \SI{100}{\micro\second}} & No & Station Receiver (jammed) \\
    Test 2.b & Kista\_driving & E1 & L1/E1 + L2/E5b & \makecell{BPSK, \SI{25}{\mega\hertz}, \SI{100}{\micro\second}} & No & Station Receiver (jammed) \\
    \addlinespace
    \midrule
    Test 3.a & Kista\_driving & L1 & L1/E1 + L2/E5b & No & \makecell{Synchronous, L1} & Station Receiver (spoofed) \\
    Test 3.b & Kista\_driving & E1 & L1/E1 + L2/E5b & No & \makecell{Synchronous, L1} & Station Receiver (spoofed) \\
    \addlinespace
    \midrule
    Test 4.a & Kista\_driving & E1 & L1/E1 + L2/E5b & No & \makecell{Synchronous,\\ L1/E1 + L2/E5b} & Station Receiver (spoofed) \\
    \bottomrule
\end{tabular}
}
\end{table*}

\begin{figure*}[]
    \captionsetup[subfloat]{farskip=2pt,captionskip=1pt}
    \setlength{\tabcolsep}{2pt} 
    \begin{tabular}{@{}p{0.5\linewidth}@{}p{0.5\linewidth}@{}}
        \centering
        \parbox{\linewidth}{%
            \centering
            \subfloat[]{
        \includegraphics[width=\linewidth]{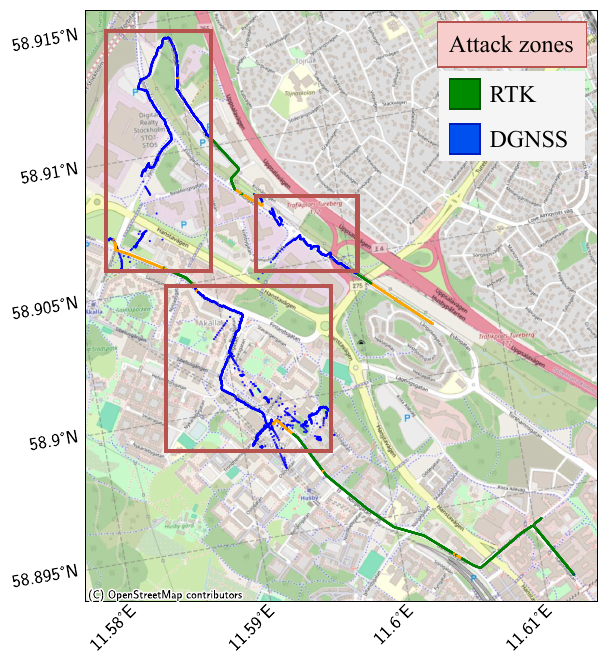}
    }
        }%
        &
        \parbox{\linewidth}{%
            \centering
            \subfloat[]{
        \includegraphics[width=\linewidth]{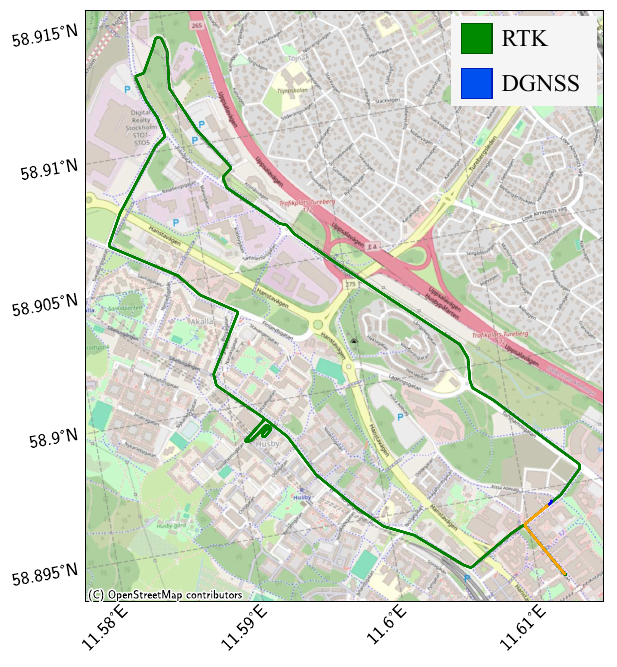}
    }%
        }%
    \end{tabular}
    \caption{\gls{rtk} solution degraded in spoofed (a) and benign (b) cases, corresponding to Test 1.c and Test 0.a, respectively.}
    \label{fig:rtkiller-result}
\end{figure*}

\textbf{Overpowered synchronous multi-constellation spoofing} -  Spoofing of the reference station successfully degraded the position solution at the rover. In \cref{fig:rtkiller-result}, the comparison between a spoofed and non-spoofed case based on \gls{rtk} solutions from RTKLib is shown. While the non-spoofed case reports a high accuracy at the rover with the large majority of the samples having fixed ambiguity and converged \gls{rtk} solution (green markers), this is not true for the spoofed case, where the rover first resorts to DGNSS (blue markers) before completely losing a valid position. This is surprising, but due to the behavior of RTKLib. As RTKLib does not change solver type based on the availability of the corrections, if corrections are unavailable, then the solution is not provided. Nevertheless, a single solution is always available as it does not depend on the corrections of the reference station. 

\begin{figure}
    \centering
    \includegraphics[width=\linewidth]{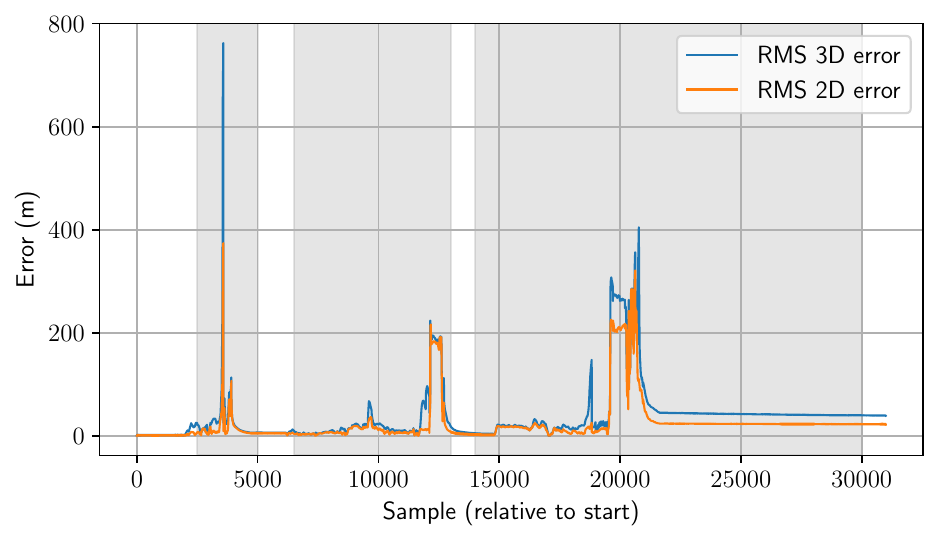}
    \caption{Root Mean Square (RMS) error of the rover PNT solution when the reference station is under spoofing. The solution is compared with the same track in a benign case.}
    \label{fig:3d-rms}
\end{figure}

The degradation is severe with an XY-RMS error of more than 50m, as shown in \cref{fig:3d-rms}. The parts of the solution affected by the adversary show that RTKLib tries to achieve a forced convergence degrading the solution status but ultimately provides an erroneous position solution. The spoofer is activated at selected periods of time, showing that the solution is quickly degraded by the adversary, but recovers almost immediately when the attacker is deactivated. 
Notably, the major contribution to the 3D RMS error at the beginning of the attack is the altitude, which immediately deviates strongly from the true value. This is of particular concern for drones and other UAVs, where sudden changes in altitude possibly lead to crashes. Nevertheless, additional sensor fusion like barometric pressure can be effectively used to counter this effect. 

\begin{figure*}[]
    \captionsetup[subfloat]{farskip=2pt,captionskip=1pt}
    \setlength{\tabcolsep}{2pt} 
    \begin{tabular}{@{}p{0.5\linewidth}@{}p{0.5\linewidth}@{}}
        \centering
        \parbox{\linewidth}{%
            \centering
            \subfloat[]{
        \includegraphics[width=\linewidth]{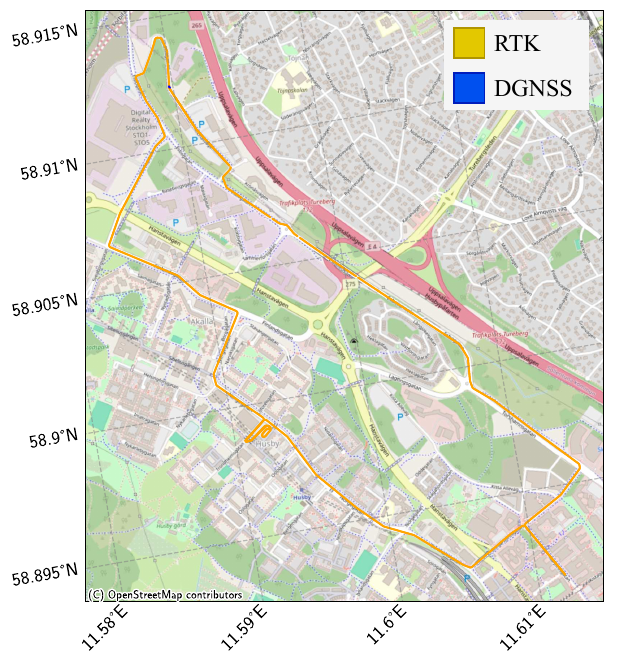}
        \label{fig:maps_test1_a}
    }
        }%
        &
        \parbox{\linewidth}{%
            \centering
            \subfloat[]{
        \includegraphics[width=\linewidth]{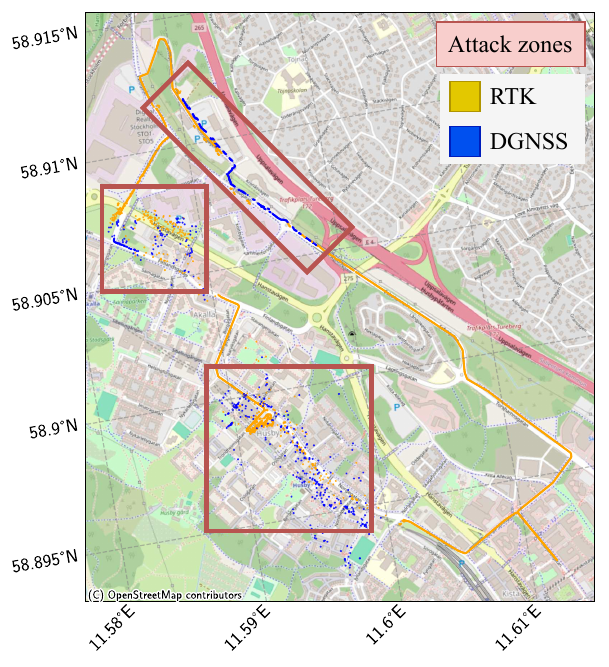}
        \label{fig:maps_test1_b}
    }%
        }%
    \end{tabular}
    \caption{Test 1.a (a) relies on legitimate corrections as reference. \gls{rtk} solution degraded by reference station spoofing only in one frequency, corresponding to Test 1.b (b). }
    \label{fig:maps_test1}
\end{figure*}

In Test 1.a the rover uses the corrections generated by the Skydel simulation software RTCM plugin, which are unspoofed high-fidelity corrections corresponding to the position of the reference station. The result is shown in \cref{fig:maps_test1_a}, with a spoofer only targeting L1/E1 already successful in modifying the rover \gls{rtk} solution \cref{fig:maps_test1_b}. In this case, the spoofer is activated and deactivated several times during the execution, showing that the rover quickly recovers from the attack, although it is immediately affected by spoofing at the reference station, as the corrections are integrated in the \gls{rtk} solution without initial check of their validity. 

\begin{figure}
    \centering
    \includegraphics[width=\linewidth]{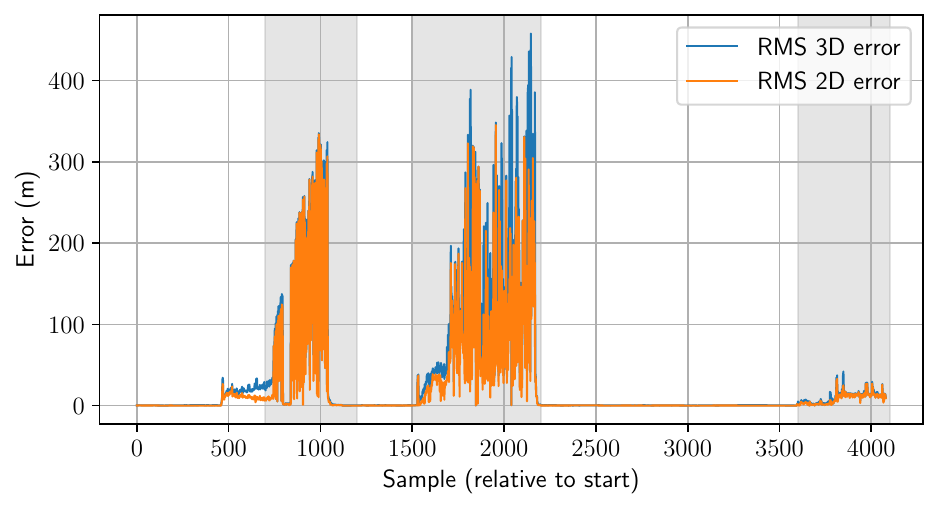}
    \caption{Root Mean Square (RMS) error of the pvt solution at the rover when the reference station is under spoofing in Test 1.b compared with the same track in Test 1.a.}
    \label{fig:3d-rms-test1}
\end{figure}

Even if the attacker only targets one frequency group, it is still successful in disrupting the rover's solution. In Test 1.c, the results are similar, but in this case, the reference station is only providing observations for one frequency band, while the rover is still configured in dual frequency mode. This suggests that RTKLib heavily relies on the available signals at the reference station and largely ignores the information already available at the rover when configured in Kinematic mode only. The error shown in \cref{fig:3d-rms-test1} shows that the error induced at the rover is considerable even if the spoofing of the reference receiver is not fully successful. 

\begin{figure}
    \centering
    \includegraphics[width=\linewidth]{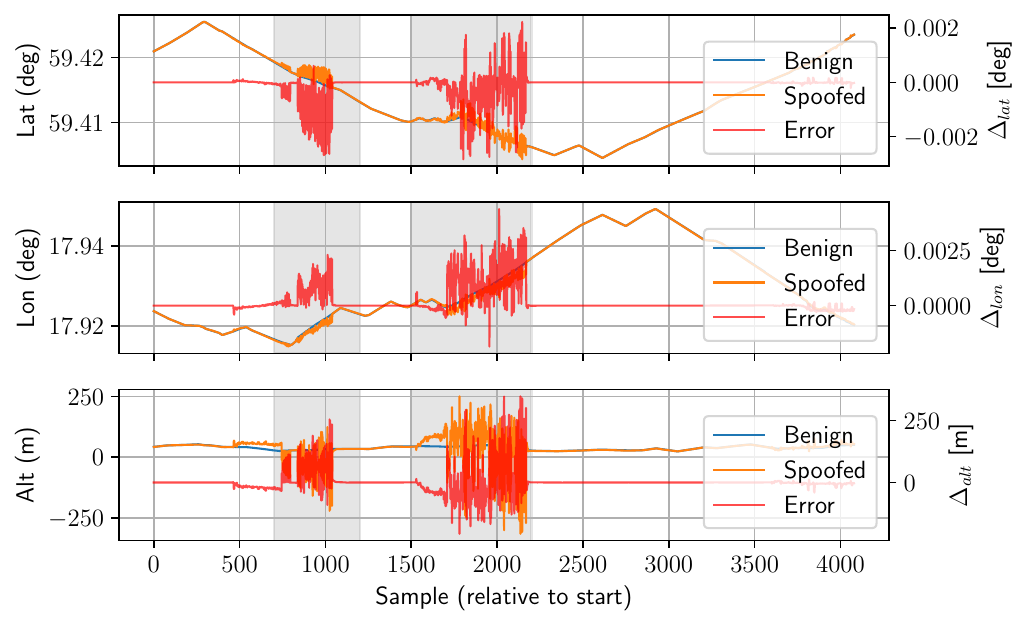}
    \caption{Deviations of the rover under spoofing in Test 1.b compared to the unaffected Test 1.a.}
    \label{fig:soldeltas_test1b}
\end{figure}

The 2D error only considers horizontal components, but the 3D error is considerably larger. This indicates that reference station spoofing is particularly problematic for UAVs, reling on GNSS for hovering. This is also clear from \cref{fig:soldeltas_test1b}, where the errors per axis are shown. The effect of the spoofer is immediately visible, with an altitude error as high as \SI{250}{\meter}. In practice, the behavior of the rover is similar to the one where it only relies on L1/E1 signals. Notably, the comparison of the ground truth and Tests 0.(a,b,c) where the receiver is completely benign, shows that the solution based only on GPS L1 is notably worse than the equivalent Galileo E1 only. 

\begin{figure*}[]
    \captionsetup[subfloat]{farskip=2pt,captionskip=1pt}
    \setlength{\tabcolsep}{2pt} 
    \begin{tabular}{@{}p{0.5\linewidth}@{}p{0.5\linewidth}@{}}
        \centering
        \parbox{\linewidth}{%
            \centering
            \subfloat[]{
        \includegraphics[width=\linewidth]{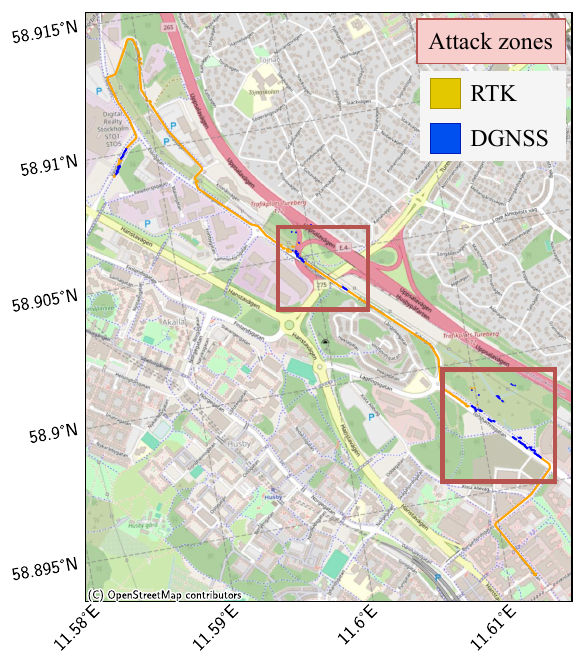}
        \label{fig:maps_test2_a}
    }
        }%
        &
        \parbox{\linewidth}{%
            \centering
            \subfloat[]{
        \includegraphics[width=\linewidth]{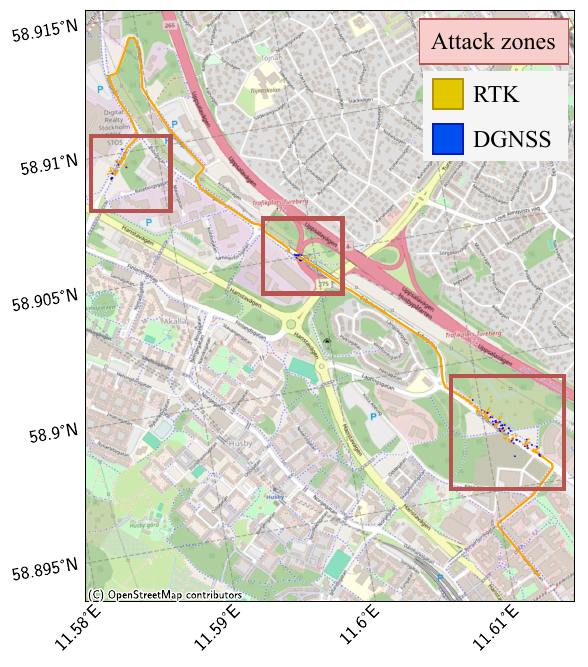}
        \label{fig:maps_test2_b}
    }%
        }%
    \end{tabular}
    \caption{\gls{rtk} solution degraded by reference station jamming in L1 frequency, corresponding to Test 2.a (a) and Test 2.b (b)}
    \label{fig:maps_test2}
\end{figure*}

\textbf{Single frequency jammer} - Test 2.(a,b) are interesting from a different perspective. \cref{fig:maps_test2} shows that jammers are very effective in disturbing the reference station. In this case, the jammer was located nearby as shown in \cref{fig:experiment-setup}, and set with a transmission power of 0dBm, equivalent to \SI{1}{\milli\watt}. The jammer has a GPS PRN-like modulation and is particularly effective against GPS receivers even at such low power, as seen in \cref{fig:maps_test2_a}. Nevertheless, despite the powerful anti-jamming filters available in the F9P front-end even the Galileo E1 solution is affected by the jammer, shown in \cref{fig:maps_test2_b} but to a lower extent. Specifically, the overall observed error is in the order of \SI{100}{\meter} for the periods affected by the jammer for GPS L1 and \SI{20}{\meter} for Galileo E1. Still, these results show that carefully designed jamming signals are very effective even against rovers that are supposedly not being affected if they depend on the corrections at the reference station. Notably, while the rover shows a lack of valid solutions in some parts of the trajectory (e.g., during the first jamming phase) the reference station always reported a valid solution, but with accuracy degraded due to jammer which ultimately led to unusable corrections.

\begin{figure*}[]
    \captionsetup[subfloat]{farskip=2pt,captionskip=1pt}
    \setlength{\tabcolsep}{2pt} 
    \begin{tabular}{@{}p{0.5\linewidth}@{}p{0.5\linewidth}@{}}
        \centering
        \parbox{\linewidth}{%
            \centering
            \subfloat[]{
        \includegraphics[width=\linewidth]{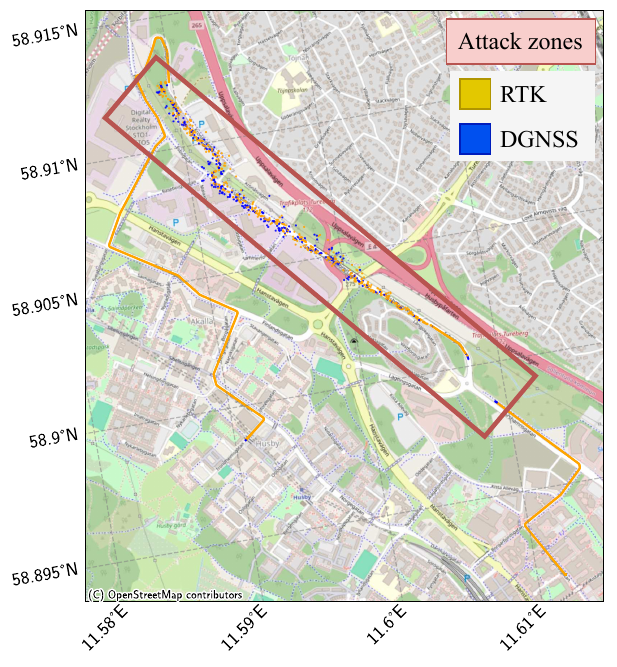}
        \label{fig:maps_test3_a}
    }
        }%
        &
        \parbox{\linewidth}{%
            \centering
            \subfloat[]{
        \includegraphics[width=\linewidth]{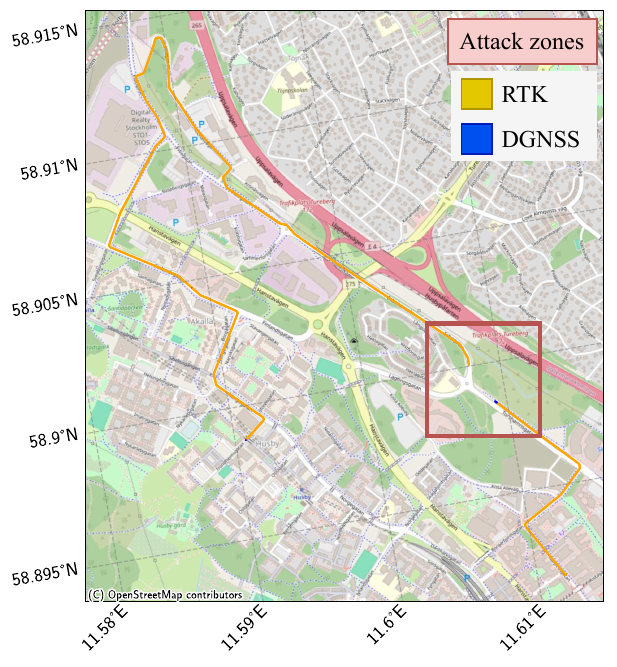}
    }%
        }%
    \end{tabular}
    \caption{\gls{rtk} solution for L1 and E1 signals, with L1 spoofing signals, corresponding to Test 3.a (a) and Test 3.b (b)}
    \label{fig:maps_test3}
\end{figure*}

\textbf{Single frequency spoofer} - Test 3.(a,b) aims to show the robustness of the \gls{rtk} solution. While the rover has a dual-frequency, multi-constellation receiver, the base station only provides corrections in the L1/E1 band. Specifically, \cref{fig:maps_test3} shows how a receiver operating in the same band as the spoofer is strongly affected by the adversary         \cref{fig:maps_test3_a}. The spoofer makes the rover solution completely unusable, although in a few epochs, the rover still produces a solution that is incorrect up to several hundred meters (3D-RMS error). On the other hand, the rover using the E1 band and Galileo signals is completely unaffected. It is important that the spoofer is seen as a jammer by the Galileo-only reference station; as a result, the solution is still slightly degraded compared to the benign case but minimally. This is due to the different modulation used by the Galileo signal. As the signal is designed to coexist with the GPS one, it will simply ignore the spoofer in L1 (this is true up to a certain level of spoofing power beyond which the front-end saturates and the spoofer effectively acts as an E1 jammer).

\begin{figure}
    \centering
    \includegraphics[width=\linewidth]{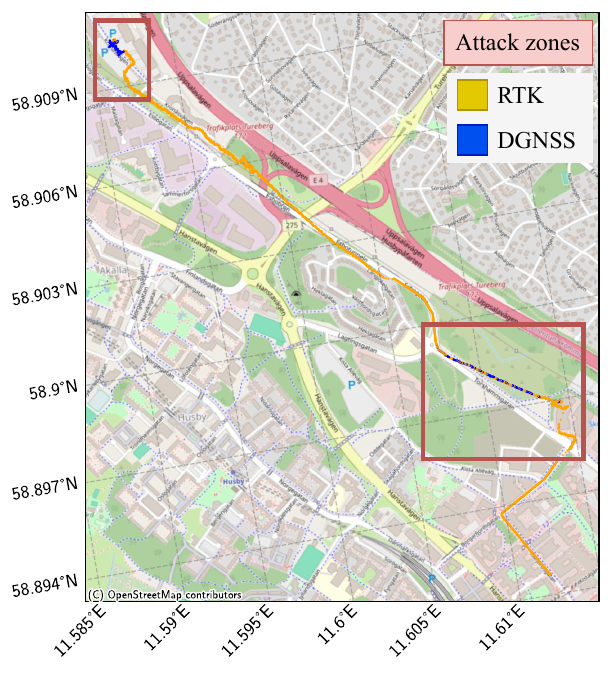}
    \caption{\gls{rtk} solution for Test 4.a: the rover is significantly affected and does not recover from the attack.}
    \label{fig:maps_test4}
\end{figure}

\begin{figure}
    \centering
    \includegraphics[width=\linewidth]{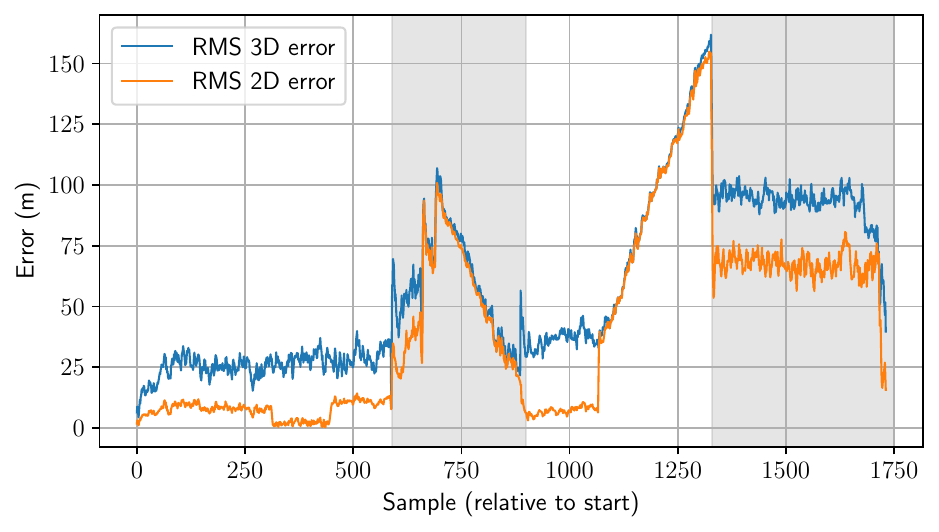}
    \caption{Root Mean Square (RMS) error of the pvt solution at the rover when the reference station is under spoofing in Test 4.(b).}
    \label{fig:3d-rms-test4}
\end{figure}

\begin{figure}
    \centering
    \includegraphics[width=\linewidth]{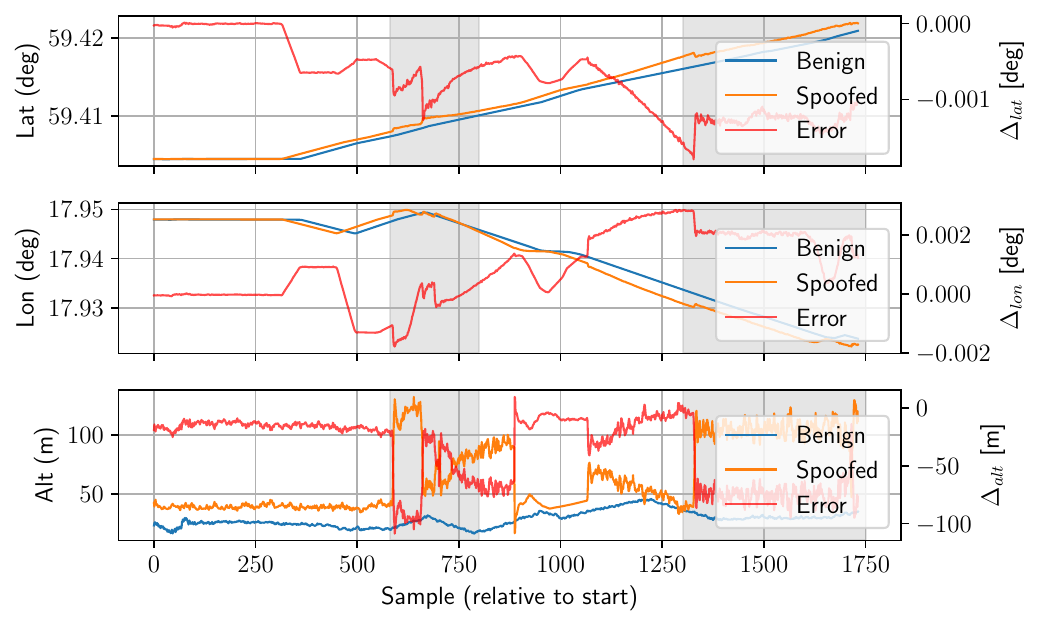}
    \caption{Deviations of the rover under spoofing in Test 4.(a). Both frequencies and constellations are under the influence of the spoofer.}
    \label{fig:soldeltas_test4a}
\end{figure}

Finally, we discuss the case of a very powerful adversary controlling synchronously both constellations and frequency bands. Such an adversary requires considerably more computational resources compared to the single constellation/frequency one butit is more likely to effectively control the victim reference station. This is shown in \cref{fig:maps_test4}: the rover corrected solution is valid and available but several hundred meters wrong compared to the true trajectory followed by the rover. This error is evident also in \cref{fig:3d-rms-test4}, particularly in the 3D calculation as altitude has the largest error component. 
Most importantly, compared to Test 3.(a,b), here the \gls{gnss} receiver does not recover quickly after the attack ends. The error persists even after the adversary is deactivated. Reasonably, this is due to the configuration of the reference station. The receiver attempts to make the spoofed solution slowly converge with the again-available legitimate one, instead of allowing a rapid change in the \gls{pnt}. While this guarantees the availability of the solution, the reference receiver provides the wrong corrections for a longer time, making the rover's recovery slower, as shown in \cref{fig:soldeltas_test4a}.

\textbf{Commercial implementation, multiconstellation and multifrequency spoofer} - Test 4.b shows the effect of spoofing on the internal \gls{rtk} engine used in a commercially available product to compute a fully resolved differential solution. In addition to the previous setup, the receiver is also configured in \textit{TIME-only} mode following survey-in, which is commonly used to establish a reference position. This mode relies on an established fixed position at the reference station obtained by subsequent observations and averaging of the calculated position. The receiver in \textit{TIME-only} mode does not report any \gls{pnt} update, but instead provides precision measurements for the NTRIP correction server. Nevertheless, the reference station can still be attacked by spoofing, with the difference that the takeover needs to be more controlled with the initial position of attacker-induced PVT on the RTK reference matching closely the actual position. 

\begin{figure}
    \centering
    \includegraphics[width=\linewidth]{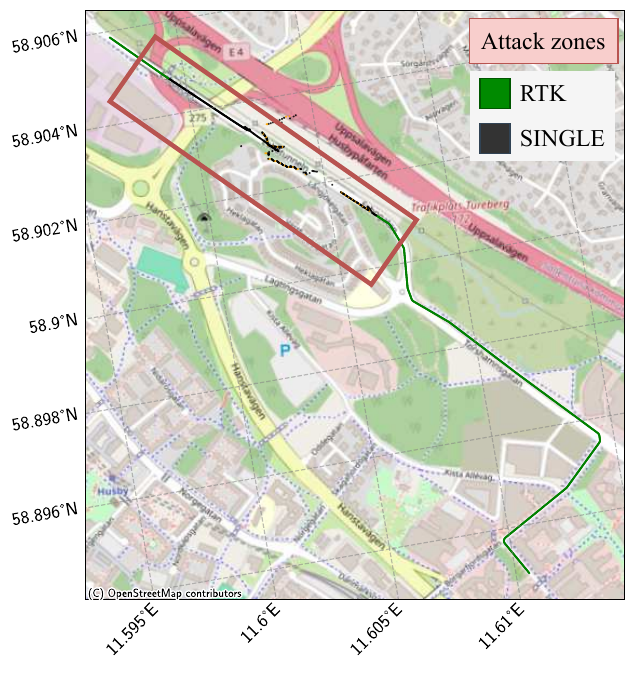}
    \caption{Ublox \gls{rtk} rover solution under attack, showing considerable deviations from the ground truth.}
    \label{fig:maps_ublox}
\end{figure}

\begin{figure}
    \centering
    \includegraphics[width=\linewidth]{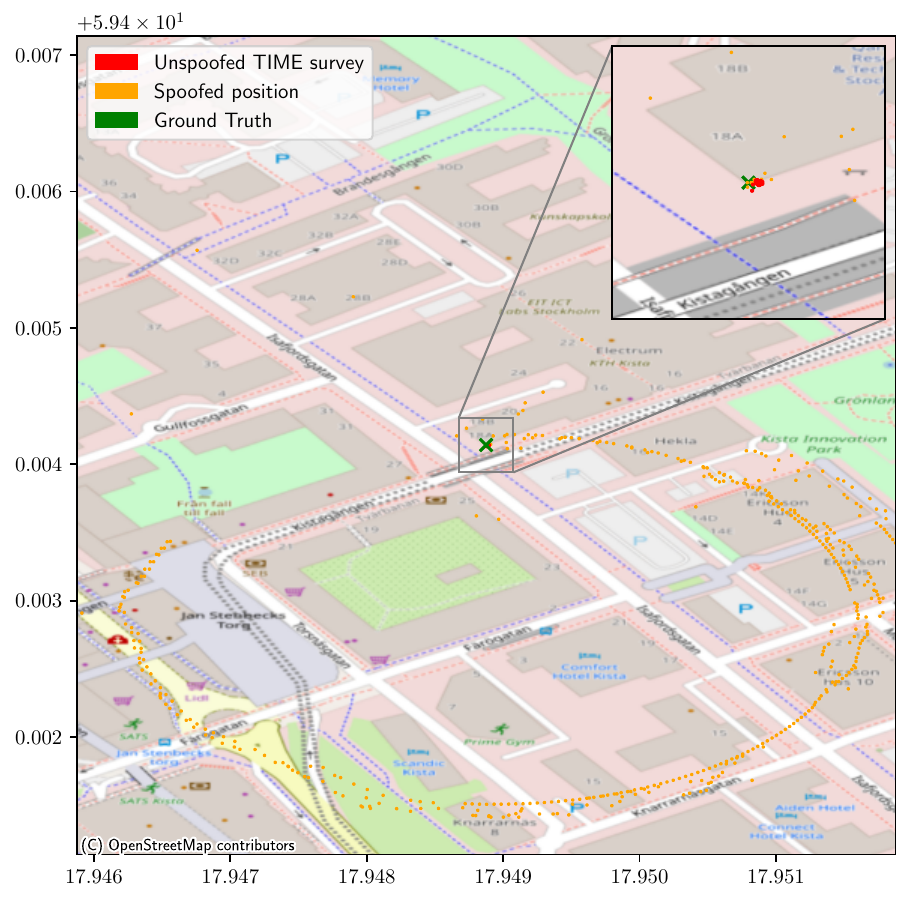}
    \caption{Ublox \gls{rtk} references station solution under attack, provided by the internal PNT engine in TIME mode and post-processing based on raw measurements.}
    \label{fig:time_ref_station}
\end{figure}

After the takeover, the reference station is successfully spoofed and the measurements provided to the rover will be valid but not referenced to the actual station position anymore and cause significant errors in the rover, as shown in \cref{fig:maps_ublox}. Compared to the RTKLib implementation, if the \gls{rtk} solution at the rover does not converge, the rover will default to its own uncorrected \gls{pnt} solution based on the local observations only and discard the correction stream. Nevertheless, the induced error is considerable (up to \SI{70}{\meter}) and constitutes a safety liability. Additionally, we observed that once the reference station completes survey-in mode, the \gls{pnt} solution is not calculated again and the position is fixed to the surveyed one. This causes an issue shown in  \cref{fig:time_ref_station}: the reference receiver reports a static position due to the successful survey-in mode, but the actual position calculated from the receiver raw measurements corresponds to the spoofed one. This behavior makes detection based upon changes in the solution hard to evaluate, given that there is no update in the \gls{pnt} solution. 

\section{Suggested countermeasures}
The tests we performed suggest that the vulnerability of the overall DGNSS/\gls{rtk} approach is not fundamentally due to the methods themselves but rather to the way the rover uses the provided corrections. This is evident in Test 1.b and Test 4.a. For Test 1.b, the rover receiver computes a correct solution based on the Galileo E1 signal, in comparison to the GPS L1. The current approach to multi-frequency multi-constellation systems is to include as many signals as possible in the calculation of the \gls{pnt} solution so that the accuracy of the measurements improves. Nevertheless, it might be beneficial in this context to adopt a more cautionary approach. The separation of solutions suggested in \cite{ZhangP:C:2019b} is a viable method in this context. Instead of only relying on corrections, the receiver should provide a solution based on their own measurements and only provide augmented measurements based not on the availability of the corrections but on their effects. Intuitively the corrected measurements should be of higher quality compared to the standalone ones at the receiver. If this is the case, the corrections are actually augmenting the receiver. As shown in \cref{sec:evaluation}, under spoofing the effect is the opposite. Practically, the suggestion is to make sure that the corrections are improving the result before applying them to the \gls{pnt} solution instead of the other way around. 

A complementary approach we propose is to detect spoofing at the reference station. General techniques used to establish a reference station include configuration of the receiver in static mode if the reference antenna location is precisely known, and survey-in techniques. The receiver should provide correction usable at the rover only in static mode while this seems not to be the case. Nevertheless, survey-in procedures can be long and complex requiring up to 24 hours of data to be accurate, making rapid establishment of reference stations problematic. Spoofers targeting the position of the reference station can be effectively thwarted using position drift monitoring and establishing an acceptance threshold for the reference station position. Knowing its antenna should not move from the established position, the concept is similar to geofencing: if the adversary forces position changes that are bigger than the allowed deviation the receiver marks the correction stream as untrusted. This is easily integrated in the existing receiver, as the instantaneous position can be derived by the continuous measurement process and compared with the survey-in result. A discrepancy between the two solutions exceeding the error level allowed by the survey-in process should trigger a new survey-in and flag an alarm. For more sophisticated adversaries spoofing the pseudoranges so that their combined use still provides the correct reference station position, detection of the attack can be performed by evaluation of residuals calculated by measured and estimated satellite position (obtained, e.g., from out-of-band ephemerides).
Furthermore, an adversary forcing a pseudorange ramp would cause a time shift at the reference station without changing its location.  Countermeasures based on clock offset and drift monitoring are well suited for this purpose, achieving both coarse and refined time-based detection of spoofing \cite{SpangheroPP:C:2023,Spanghero2022}. 

Furthermore, the rover can rely on several reference stations when available, obtaining correction streams from multiple stations spread over space, so that they are not all spoofed/jammed yet close enough to the rover to provide relevant corrections and compare the corrected solution. If certain correction providers do not match with the suggested result from the majority of the reference stations, that provider can be voted out and ignored in future iterations of the differential GNSS solver. This approach is often used in advanced receivers to refine the corrections obtained by the network, but there is no constraint to using it as a security hardening method.

\section{Conclusions}
\label{sec:conclusion}

We presented a structured analysis of the vulnerabilities of \gls{rtk} to spoofing at the reference station. We show, based on attack simulation with the Safran Skydel RF level simulator, that controlling the reference station is not only possible but also practical given its static and precisely known position. This is true for simple jamming, asynchronous spoofing, synchronous signal lift-off, and even for self-established base stations via survey-in. We showed how the current open-source implementation of \gls{rtk} is vulnerable to adversarial control and that targeted adversarial action affects all rovers connected to the reference station under spoofing. Specifically, it affects the quality of the DGNSS/\gls{rtk} solution and the solution accuracy at the receiver. Additionally, we propose validation methods for the corrections based on both measurements available at the rover and state estimation at the reference station itself. 

While L-band and PPP corrections could supersede the need to use fixed reference stations, the latter will still be important for high-accuracy services. The addition of new features and signals will raise the bar for the spoofer to obtain control of the reference station, but still, would harm the integrity of its \gls{pnt} solution. Readily deployable protection methods can be used to harden the reference station infrastructure and augment the rover processing of corrections to make them more reliable. 

\section*{Acknowledgements}
The NSS group is part of Safran Minerva Academic program and the Skydel software was granted through the academic partnership, including the additional plugin licenses that made this work possible. This work was supported in parts by the national strategic research area on security and emergency preparedness.

\bibliographystyle{IEEEtran}
\bibliography{sample_full}



\end{document}

%% file: acronyms.tex
\newacronym{scd}{SCD}{Spectral Correlation Density}
\newacronym{cw}{CW}{Continuous Wave}
\newacronym{fam}{FAM}{Fourier Accumulation Method}
\newacronym{snr}{SNR}{Signal to Noise Ratio}
\newacronym{bpsk}{BPSK}{Binary Phase Shift Key}

\newacronym{ppd}{PPD}{Privacy Protection Device}
\newacronym{tls}{TLS}{Transport Layer Security}

\newacronym{gnss}{GNSS}{Global Navigation Satellite System}
\newacronym{gps}{GPS}{Global Positioning System}
\newacronym{nmea}{NMEA}{National Marine Electronics Association}
\newacronym{nasa}{NASA}{National Aeronautics and Space Administration}
\newacronym{agps}{A-GPS}{Assisted/Augmented GPS}
\newacronym{scer}{SCER}{Security Code Estimation and Replay}
\newacronym{nma}{OS-NMA}{Open Service Navigation Message Authentication}
\newacronym{sis}{SIS}{Signal In Space}
\newacronym{raim}{RAIM}{Receiver autonomous integrity monitoring}
\newacronym{pnt}{PNT}{Position-Navigation-Time}
\newacronym{rtk}{RTK}{Real-time Kinematics}
\newacronym{agc}{AGC}{Automatic Gain Control}
\newacronym{meo}{MEO}{Medium Earth Orbit}
\newacronym{rssi}{RSSI}{Received Signal Strength Indicator}
\newacronym{llh}{LLH}{Latitude, Longitude and Height}
\newacronym{osnma}{OSNMA}{Open Signal Navigation Message Authentication}
\newacronym{sqm}{SQM}{Signal Quality Monitoring}
\newacronym{los}{LOS}{Line of Sight}
\newacronym{ntrip}{NTRIP}{Networked Transport of RTCM via Internet Protocol}

\newacronym{nco}{NCO}{Numerically Controlled Oscillator}
\newacronym{pll}{PLL}{Phase-Locked Loop}
\newacronym{ecef}{ECEF}{Earth Centered Earth Fixed}
\newacronym{pca}{PCA}{Principal Components Analysis}
\newacronym{mle}{MLE}{Maximum Likelyhood Estimator}

\newacronym{sdr}{SDR}{Software Defined Radio}
\newacronym{sdrs}{SDR}{Software Defined Radios}
\newacronym{ots}{OTS}{Off-the-Shelf}
\newacronym{rt}{RT}{Real-Time}
\newacronym{ide}{IDE}{Integrated Development Environment}
\newacronym{fifo}{FIFO}{First-In-First-Out}
\newacronym{sma}{SMA}{SubMiniature Version A}
\newacronym{rf}{RF}{Radio Frequency}
\newacronym{uav}{UAV}{Unmanned Aerial Vehicle}
\newacronym{vtol}{VTOL}{Vertical Take Off and Landing}
\newacronym{ota}{OTA}{Over-The-Air}
\newacronym{ism}{ISM}{Industrial Scientific and Medical}

\newacronym{swap}{SWaP}{Size, Weight and Power}
\newacronym{imu}{IMU}{Inertial Measurement Unit}
\newacronym{cots}{COTS}{Commercial Off the Shelf}
\newacronym{som}{SoM}{System on Module}